\documentclass[twocolumn,eqsecnum,aps,prd,showpacs,preprintnumbers,amsmath,amssymb,nofootinbib,floatfix]{iopart}

\usepackage{bm}
\usepackage{graphicx,epsf}
\usepackage{amssymb}

\newcommand\fh{\hat{f}}

\newcommand\hh{\hat{h}}

\newcommand\sh{\hat{\sigma}}

\newcommand\ph{\hat{\phi}}

\newcommand\gsim{\gtrsim}

\renewcommand\({\left(}

\renewcommand\){\right)}

\renewcommand\[{\left[}

\renewcommand\]{\right]}

\newcommand\eq[1]{Eq.~(\ref{#1})}

\newcommand\ee{\end{equation}}

\newcommand\be{\begin{equation}}

\newcommand\eea{\end{eqnarray}}

\newcommand\bea{\begin{eqnarray}}

\def\calp{{\cal P}}

\newcommand\sub[1]{_{\rm #1}}

\newcommand\half{^{1/2}}

\newcommand\fnl{{f\sub{NL}}}
\newcommand\tnl{{\tau\sub{NL}}}

\newcommand\zetai{\zeta\sub{inf}}
\newcommand\zetas{\zeta_\sigma}

\usepackage{pstricks}

\begin{document}
    \title{Curvature perturbation from symmetry breaking the end of inflation}

    \author{Laila Alabidi and David Lyth\\ Physics Department, Lancaster University,LA1 4YB}

    \begin{abstract}
    We consider a two-field hybrid inflation model, in which the
    curvature perturbation is predominantly generated at the end
    of inflation. By finely tuning the coupling of the fields to the waterfall
    we find that we can get a measurable amount of
    non-gaussianity.
    \end{abstract}
      \maketitle

    \section{Introduction}

\begin{figure}
        \begin{center}
            \begin{pspicture}(-4,-4)(4,4)
                \rput[l]{90}(-3.5,){INFLATION}
                \psline[linestyle=dashed,linewidth=0.2](-3,3.5)(3,3.5)
                \psline[arrows=->](-2,3.5)(-2,2)
                \rput[l](-2.5,2.5){$t_1$}
                \psline[arrows=->](0,3.5)(0,-2)
                \rput[r](0.5,2.5){$t_2$}
                \psline(-3,0)(3,0)
                \pscurve[linestyle=dashed](-3,0)(-2,2)(0,-2)(2,2)(3,0)
                \rput[r](-1.3,1){$\delta{}N\sim 0$}
                \rput[l](1.2,1){$\delta{}N\sim 0$}
                \rput*(0,-1){$\delta{}N=N_e$}
                \rput(0,4){Horizon Exit}
                \rput[r]{90}(-3.5,-0.5){NON INFLATION}
                \rput(0,-3.5){surface at the}
                \rput(0,-4){end of inflation}
                \psline[arrows=->](-0.5,-3.4)(-0.3,-2)
                \rput(3,-1){surface of constant}
                \rput(3,-1.5){energy density}
                \psline[arrows=->](3,-1)(2.5,0)
            \end{pspicture}
            \caption{Single field inflation ends on a surface of constant energy density, we however are considering the case of multiple-fields
            driving inflation, and thus inflation ends on a surface of non-constant energy density, as represented by the dotted line.
            It is clear to see that $t_1$ and $t_2$
            produce different amount of inflation, thus the family of trajectories is still curved at the end, and the curvature perturbation is increased
            by an amount $N_e=\delta{}N$. }
        \end{center}
    \label{ex2}
    \end{figure}
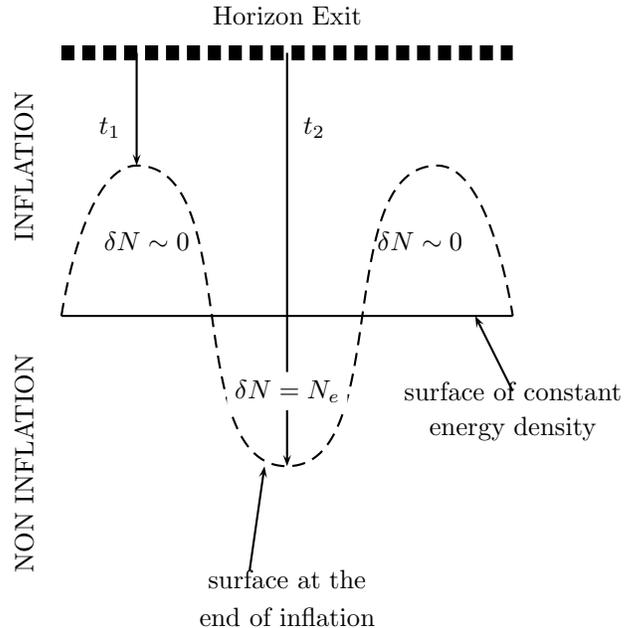

      We consider the scenario \cite{first,lyth,salem}
      in    which a possibly dominant contribution to
the curvature perturbation is generated at the end of inflation.
As illustrated in Figure \ref{ex2}, this contribution is generated
if the end of  inflation (taken to be abrupt) occurs on a spacetime
slice which is not one of uniform energy density.

This mechanism  can operate most  cleanly in hybrid inflation, where slow-roll
is valid right up to the end of inflation so that slices of uniform density
corresponds to slices of equal potential, or in other words of uniform
inflaton field. Inflation in that case ends when
the mass-squared of the waterfall field falls through zero. The mechanism
operates if the mass-squared depends not only on the inflaton, but also
on some other light field. The general formalism
for this scenario was worked out in
\cite{lyth}.

As it stands, this mechanism may be deemed unattractive because the
introduction of the extra light field
seems ad hoc, designed only to produce the desired
effect. In this paper we propose that inflation actual takes place in the
space of a two-component object
$(\phi,\sigma)$.  During inflation the potential depends only on
$\phi^2+\sigma^2$, corresponding to an $SO(2)$ symmetry. But as illustrated
in Figure \ref{model} the end of inflation occurs on an ellipse, which breaks
the symmetry and generates a contribution to the curvature perturbation at the
end of inflation. We choose inverted hybrid inflation \cite{inverted}
instead of the original model \cite{originalhybrid} so that we can reproduce
the observed negative spectral tilt \cite{wmap3}.

    \section{The  Model}

The potential is
    \bea\label{pot}
    \lefteqn{V(\phi,\sigma,\psi)=}\nonumber\\
    &&{ }V_0-\frac{1}{2}m^2\left(\phi^2+\sigma^2\right)-
\left(f\phi^2+g\phi\sigma+h\sigma^2\right)\frac{\psi^2}{2}\nonumber\\
    &&{ }+\frac{m_{\psi}^2}{2}\psi^2
    \eea
The waterfall field
$\psi$ is held at $0$ during inflation and rolls
    rapidly to
    it's true minimum when
    \begin{equation}\label{cond}
    m_{\psi}^2=f\phi_e^2+g\phi_e\sigma+h\sigma^2
,
    \end{equation}
where $\phi_e(\sigma)$ is the value of $\phi$ at the end
    of inflation.

     \begin{figure}
        \begin{center}
        \begin{pspicture}(-4,-4)(4,4)
        \psline[arrows=<->](-4,0)(4,0)
        \psline[arrows=<->](0,-4)(0,4)
        \rput{45}(0,0){\psellipse(0,0)(3.5,1.5)}
        \psellipse[linestyle=dotted](0,0)(2,2)
        \psellipse[linestyle=dotted](0,0)(3.5,3.5)
        \psline[arrows=->,linewidth=0.2](0,0)(0,2)
        \rput(1.5,1){Inflaton trajectory}
        \rput(4,0.2){$\sigma$}
        \rput(0.2,4){$\phi$}
        \end{pspicture}
        \caption{Representation of the setup, where the surface at the end of inflation is an ellipse, therefore varying the
        trajectory (or the orientation of the ellipse)
        varies the amount of inflation. The dotted circles correspond to scenarios in which all trajectories are equal,as in single field inflation}
        \end{center}
        \label{model}
    \end{figure}
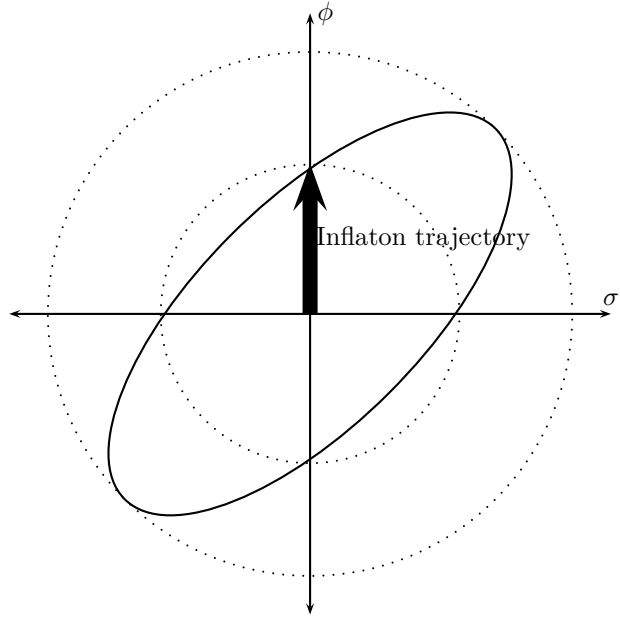

    During inflation, we take the trajectory
    to be along the $\sigma=0$ path, and the potential behaves as a quadratic one $V=V_0-m^2\phi^2/2$. We allow the elliptic surface of the end of inflation to have any orientation
    (see figure\ref{model}), which is equivalent to fixing the ellipse and varying the trajectory.
    This is accounted for in the cross coupling term in the potential
    $\left(-g\phi\sigma\psi^2/2\right)$.
    For such a setup, either all
    slow-roll conditions are satisfied till $\psi$ is
    destabilized, or the violation of one 'pauses' inflation
    until the amplitude of the perturbations decays and inflation
    resumes. In this paper we consider the former. To analyse this
    case, we will use the equations derived in \cite{lyth}.

    \section{The Equations}

    The curvature perturbation is $\zeta=\zetai+\zetas$. The first term
    is the constant contribution of $\delta\phi$, and the second term is the
    constant
    contribution of $\delta\sigma$ which is generated only at the end of inflation.
    From \cite{lr,lyth},
    \begin{equation}\label{curv}
    \zetas=N_e'\phi_e'\delta\sigma+\frac{1}{2}\left[2N_e''(\phi_e')^2+N_e'\phi_e''\right](\delta\sigma)^2
    \end{equation}
    where:
    \begin{equation}\label{ne}
    N_e'=\frac{dN_e}{d\phi_e}=\frac{1}{\sqrt{2\epsilon_e}}
    \end{equation}

    \begin{equation}\label{nee}
    N_e''=\frac{d^2N_e}{d\phi_e^2}=\frac{2\epsilon_e-\eta}{2\epsilon_e}
    \end{equation}

    The slow roll parameters are:
    \be\label{eta}
    \eta=-\frac{m^2}{V_0}  \\
    \ee
    \be\label{eps}
    \epsilon_e=\frac{m^4}{2V_0^2}\phi_e^2=\eta^2\frac{m_{\psi}^2}{2f}
    \ee

    and the spectral index for this model is:

    \be
        n-1=2\eta
    \ee

    {}From (\ref{cond}), for $\sigma=0$ we have
    $\phi_e^2=m_{\psi}^2/f$.
    Thus:
    \be\label{pe}
    \phi_e'=-\frac{g\phi_e+2h\sigma}{2f\phi_e+g\sigma}=-\frac{g}{2f}
    \ee
    and
    \bea\label{pee}
    \phi_e''&=&-\frac{g\phi'_e+2h}{2f\phi_e+g\sigma}+\frac{g\phi_e+2h\sigma}{(2f\phi_e+g\sigma)^2}(2f\phi'_e+g)\nonumber\\
    & = & {}\frac{g}{2m_{\psi}\sqrt{f}}\left(\frac{g}{2f}-\frac{2h}{g}\right)
    \eea

    Substituting equations
    (\ref{ne}$\rightarrow$ \ref{pee}) in equation (\ref{curv}) we get:
    \bea
    \zetas&=& -\frac{g}{2f\sqrt{2\epsilon_e}}\delta\sigma\nonumber\\
    &+&\frac{1}{2}\left[\left(2-\frac{\eta}{\epsilon_e}\right)
    \frac{g^2}{4f^2}+\frac{g}{2m_\psi\sqrt{2\epsilon_ef}}\left(\frac{g}{2f}-\frac{2h}{g}\right)\right](\delta\sigma)^2 \label{38}
    \eea

    \section{Parameter Constraints I}

    We analyse first the
case $\calp_{\zetas}\gg \calp_{\zetai}$, where
    $\zetai$ is the inflaton contribution to $\zeta$ and $\calp$ are the
spectra.  Then $\calp_\zeta\simeq \calp_{\zetas}$. To
agree with the
    observation $\zetas$ must be  almost gaussian, so
    the first term of \eq{38} must dominate giving
    \be
    \mathcal{P}_{\zetas}=\frac{g^2}{4f^2}\left(\frac{1}{2\epsilon_e}\right)\left(\frac{H_k}{2\pi}\right)^2
    \ee
    As $\mathcal{P}_{\zetai}=
\frac{1}{2\epsilon}\left(\frac{H_k}{2\pi}\right)^2$ thus
     $\mathcal{P}_{\zetas}=\frac{g^2}{4f^2}\frac{\epsilon}{\epsilon_{e}}
\mathcal{P}_{\zetai}$. From slow roll we have
    $3H\dot{\phi}=-\frac{dV}{d\phi}$, and $N=H\Delta{}t$, thus
    $\frac{\phi_k}{\phi_e}=\exp{(\eta{}N)}\sim 1$. Since
    $\epsilon/\epsilon_e=\left(\phi_k/\phi_e\right)^2$, $\mathcal{P}_{\zetas}$
    simplifies to:
    \be\label{spec}
        \mathcal{P}_{\zetas}=\frac{g^2}{4f^2}\mathcal{P}_{\zetai}
    \ee

Since the slow-roll parameters in this model satisfy
$\epsilon\ll \eta$, the spectral index  is $n=1+2\eta$ whether or not
$\zeta_\sigma$ dominates. We can choose $\eta$ so that $n=0.95$ in agreement
with observation.

Non-gaussianity  is specified by
 parameters $\fnl$
and $\tnl$, defining  respectively the bispectrum and
trispectrum of $\zeta$. We write \be \zetas = \zeta_g + x
\zeta_g^2 , \ee where $\zeta_g =N\sub e'\phi\sub e \delta \sigma$
is practically gaussian. Then

\bea
(3/5)\fnl &=&  x  \\
&=&
 \frac{\eta{}f}{g}\left(\frac{2h}{g}-\frac{g}{2f}\right)
   ,
   \label{fnl}
\eea
 and $\tnl= 4x^2$.
For $\fnl$ to be observable we need $h/g\gg 1$, which means that
$\fnl$ is positive while $\tnl$ is bounded to be $\lesssim{}10^4$
\cite{KK}. Its observational bound is $-54< \fnl < 114$
\cite{wmap3}.

\begin{figure}
    \centering\includegraphics[angle=270, width=\linewidth, totalheight=2.5in]{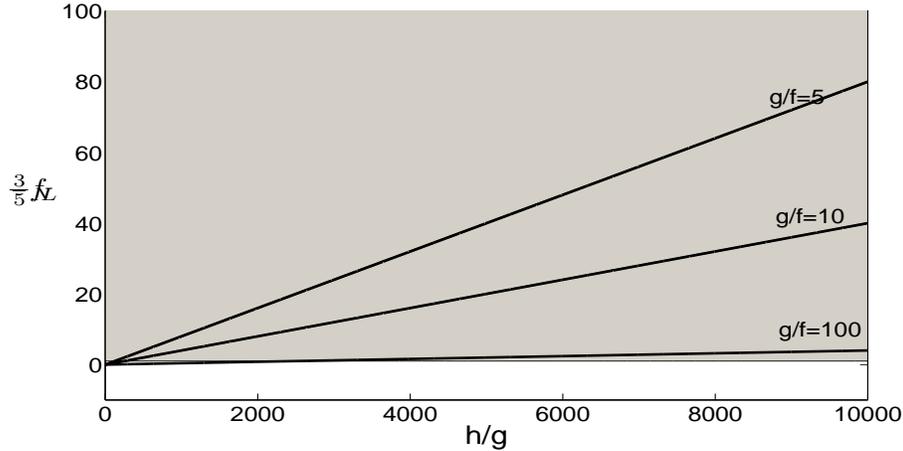}
    \caption{Plot of $\frac{3}{5}f_{NL}$ vs. $h/g$ for three values of $g/f$. It is clear from the graph
     that we have a measurable (significant) non-gaussianity for $h\gg{}g$. The shaded region corresponds to the region which \emph{can} be measured, and which has not yet been ruled out}
    \label{result}
    \end{figure}

In Figure \ref{result} we plot $f_{NL}$ versus $h/g$, taking
 $g/f=10$ and $2\eta=0.05$.
 The interesting result is that the multi-field
    hybrid model can generate a significant amount of
    non-gaussianity through the termination of inflation.
Note    that for $h>500g$ we get $f_{NL}\gsim 1$, which is the
    condition for eventual detectability.

Now we come to the case $\calp_{\zetas} \ll \calp_{\zetai}$. Then
\be \zeta \simeq \zetai + \delta\hat\sigma^2 , \ee with \be \delta
\hat \sigma^2 \simeq
\frac{g}{2m_\psi\sqrt{2f\epsilon_e}}\left(\frac{g}{2f}-\frac{2h}{g}\right)\delta\sigma^2
. \ee The non-gaussian fraction is \be r\sub{ng} \equiv \(
\frac{\calp_{\delta \hat\sigma}}{\calp_{\zetai}} \)\half \simeq \[
\frac{g}{2m_\psi
\sqrt{2\epsilon_ef}}\left(\frac{g}{2f}-\frac{2h}{g}\right) \]\half
, \ee which is related to the non-gaussianity parameters by \be
r\sub{ng} \sim \( |\fnl| \calp_\zeta^{1/2} \)^\frac13 \simeq   \(
\tnl \calp_\zeta \)^\frac14 \label{rngexp} . \ee The observational
bound on $\fnl$ requires $r\sub{ng}< 0.2$ while the bound on
$\tnl$ requires $r\sub{ng}< 0.07$ \cite{lyth2}.

    \section{Parameter Constraints II}

    Now we fix the orientation of the  ellipse, and allow a variable unperturbed
    trajectory making an angle $\theta$ with the $x$-axis. If $\theta$
    is regarded as a parameter of the theory this is equivalent to
    the previous formulation. But if inflation has lasted for a long
    time before the observable Universe leaves the horizon, the
    observable Universe will be surrounded by a huge inflated patch,
    and for a random location of the observable Universe $\theta$
    will have a  flat probability distribution.

    We convert to a coordinate system in which the ellipse is
    aligned with the axes, i.e. where the ellipse equation is
    given by:

        \be\label{form1}
            \fh\ph^2+\hh\sh^2=1
        \ee

    and $f,g,h$ are related to $\fh,\hh$ by:

        \bea\label{rel1}
            f&=&\fh\cos^2\theta+\hh\sin^2\theta\nonumber\\
            g&=&2\sin\theta\cos\theta(\hh-\fh)\nonumber\\
            h&=&\fh\sin^2\theta+\hh\cos^2\theta
        \eea

    In this new coordinate system the spectrum
    equation(\ref{spec}) is given by:

        \be
            \calp_{\zetas}=\left(\frac{X}{\cot\theta+(X+1)\tan\theta}\right)^2\calp_{\zetai}
        \ee

    where $\theta$ is the angle between the new and old coordinate
    systems, and $X=\frac{\hh}{\fh}-1$ is the symmetry breaking term. Since we require $\calp_{\zetas}>\calp_{\zetai}$
    then $X>\frac{\tan\theta+\cot\theta}{1-\tan\theta}$ which translates to a constraint on the range of
    $\theta$:

        \be
        \frac{X-\sqrt{X^2-4X-4}}{2(X+1)}<\tan\theta<\frac{X+\sqrt{X^2-4X-4}}{2(X+1)}
        \label{range}
        \ee

\begin{figure}
        \centering\includegraphics[angle=270,width=\linewidth,totalheight=2.5in]{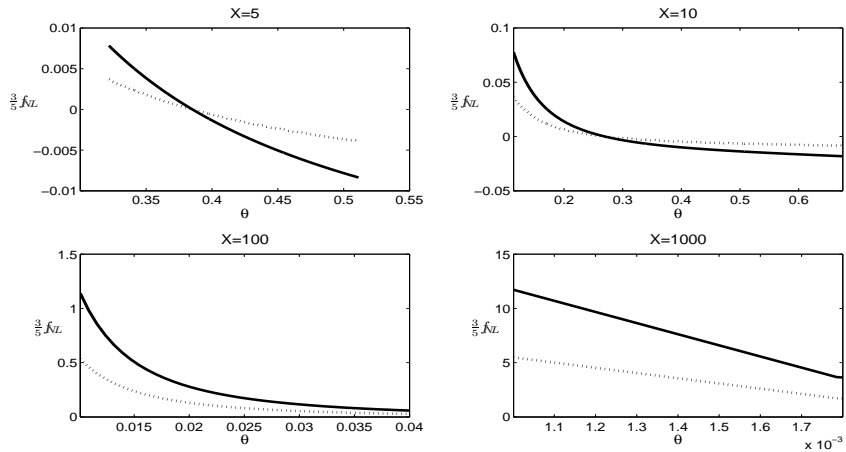}
        \caption{$\frac{3}{5}f_{NL}$ versus the angle $\theta$ for various degrees of symmetry breaking the end of inflation.
        The bold lines correspond to the
        WMAP upper bound at $1\sigma$ $n=0.953$ and the dotted lines corresponds to the $2\sigma$ upper bound $n=0.978$.}
        \label{fnl_theta}
    \end{figure}

    The non-gaussianity from end of inflation \ref{fnl} becomes:

        \bea\label{fnl_new}
        \frac{3}{5}f_{NL}&=&-\frac{2\eta}{X^2\sin^2(2\theta)}\Bigg\{-\frac{X}{4}\sin(2\theta)(X+2)+X+1\nonumber\\
                        &&+\frac{X^2}{4}\sin(2\theta)(\cos(2\theta)-\sin(2\theta))\Bigg\}
        \eea

    which we plot in figure \ref{fnl_theta} for various
    degrees of symmetry breaking.

    \section{Discussion}

    Judging from figure \ref{fnl_theta}, we note that for a
    measurable $f_{NL}$ in the case where
    $\calp_{\zetas}>\calp_{\zetai}$, we require a high degree of
    symmetry breaking the end of inflation $X\gg{}1$ and a small
    angle of orientation $\theta\ll{}\pi/4$. If inflation lasted for a long time before the observable Universe left the
    horizon, such a small angle requires the observable Universe to be
    in a special location. Anthropic selection cannot be invoked to
    favour that location, because it  applies only to the
    normalization $\calp_\zeta$ of the spectrum \cite{lyth2}. We
    conclude that within this model, an observable value of $\fnl$ is
    allowed but not favoured.

\end{document}